\documentclass[aps,pra,twocolumn,showpacs,superscriptaddress,english,graphics]{revtex4}
\usepackage{graphicx,color}
\usepackage{amssymb,amsmath,amsfonts,mathrsfs,eufrak,fancyhdr}
\usepackage{amsfonts}
\usepackage{bm}

\def\<{\langle}
\def\>{\rangle}

\def\set#1{{\sf #1}}

\def\hil#1{{\mathscr{#1}}}

\def\Tr{\operatorname{Tr}}

\def\cU{{\mathcal U}}
\def\cH{{\mathcal H}}

\def\avg#1{\langle #1 \rangle}

\def\ket#1{| #1 \rangle}
\def\braket#1#2{\langle #1|#2 \rangle}
\def\Tr{\mathrm{Tr} \, }

\usepackage{epstopdf}

\begin{document}
\title{Relaxation to Equilibrium in a Quantum Network }
\author{J.\ Novotn\'y}
\affiliation{Department of Physics, Czech Technical University in Prague,
115 19 Praha 1 - Star\'e M\v{e}sto, Czech Republic}
\author{A.\ Mariano}
\affiliation{ENEA, Italian National Agency for New Technologies, Energy, and Sustainable Economic Development,
via G. Petroni 15/F, I-70124 Bari, Italy}
\author{S.\ Pascazio}
\affiliation{Dipartimento di Fisica and MECENAS, Universit\`a di Bari, I-70126 Bari, Italy}
\affiliation{INFN, Sezione di Bari, I-70126 Bari, Italy}
\author{A.\ Scardicchio}
\affiliation{The Abdus Salam International Center for Theoretical Physics, Strada  Costiera  11,  34151  Trieste,  Italy}
\affiliation{INFN Sezione di Trieste, I-34127 Trieste, Italy}
\author{I.\ Jex}
\affiliation{Department of Physics, Czech Technical University in Prague,
115 19 Praha 1 - Star\'e M\v{e}sto, Czech Republic}

\begin{abstract}
The approach to equilibrium of quantum mechanical systems is a topic as old as quantum mechanics itself, but has recently seen a surge of interest due to applications in quantum technologies, including, but not limited to, quantum computation and sensing. The mechanisms by which a quantum system approaches its long-time, limiting stationary state are fascinating and, sometimes, quite different from their classical counterparts. In this respect, quantum networks represent a mesoscopic quantum systems of interest. In such a case, the graph encodes the elementary quantum systems (say qubits) at its vertices, while the links define the interactions between them. We study here the relaxation to equilibrium for a fully connected quantum network with CNOT gates representing the interaction between the constituting qubits. We give a number of results for the equilibration in these systems, including analytic estimates. The results are checked using numerical methods for systems with up to 15-16 qubits. It is emphasized in which way the size of the network controls the convergency. 
\end{abstract}


\maketitle

\section{Introduction}

Quantum networks \cite{Kimble,Yurke} find a wide range of applications in
quantum theory and information processing. In rather general terms, a
quantum network is an ensemble of quantum systems--typically
qubits--with a prescribed set of interactions between them, defining
the overall pattern that enables them to carry out specific tasks.
Quantum networks can have different degrees of complexity and hence
also execute tasks that can be more or less sophisticated. Quantum
networks can be used to carry out computations, communications or
storage of quantum information \cite{Mahler, Nielsen}.

The information about the details of the network and the mutual
interactions between its constituting parts is efficiently encoded
into graphs. The vertices represent the quantum systems and the
edges (links) the interactions between the network elements. In the
simplest case the links of the graph are static and unchangeable. In
such a case we assume that the network dynamics is described by a unitary
dynamics which is not changing in time. However, it is not difficult
to generalize such a structure to encompass more general situations.
The links between the elements of the network can be activated or
terminated and the underlying graph encodes then only the
potentiality of two or more elements to interact. Such situations
can describe, for instance, a quantum gas where the elements of the
network are not qubits but atoms or molecules and by using the
concept of network we follow the formation of the
asymptotic-stationary states due to elementary interactions between
them \cite{Maxwell, Boltzmann}. In such a case each link-edge is given a weight representing
the probability with which a given interaction is carried out and
hence we follow the evolution of the system with sufficient time
resolution. Such a situation is inherently random and while the
elementary time evolution (represented by a given sequence of operations) is unitary and given by the product of individual unitary operations, the overall evolution is
non-unitary \cite{Holevo}. Even though all input states are
available, the evolution of the system tends in general to an
attractor space of the network and does not take place in a
subspace of the original Hilbert space. The evolution of the system
is described by the repeated application of a completely positive map. The
basic task in solving the dynamics of the network is twofold: one
first determines the asymptotic space, and then finds the rate at
which the system approaches this subspace. Such a task is in general
intimately linked to the choice of the graph chosen, the weight of
the links and naturally the form of interaction between the
constituting parts. In the following we will focus on qubit networks
with CNOT operations between any chosen pair of qubits, hence the
underlying graph will be the fully connected graph \cite{NAJ,NAJ1}.

These collision models are reminiscent of the popular Boltzmann gas model of statistical physics \cite{Bgas,Bgas2,Maxwell}, in which one has sufficient time resolution to guarantee that only bipartite interactions be considered.
Models of this type are routinely used in the study of the approach to equilibrium, transport phenomena, decohence and dephasing, and the study of temporal synchronization \cite{Palma2017}. In this context, it is worth of notice that the approach to equilibrium, and lack thereof, of a quantum system has received a lot of attention due to both theoretical advances \cite{eth1,eth2,eth3, mbl1,mbl2,mbl3,mbl4,mbl5,mbl6,mbl7} and experimental results \cite{al1,mbl8}. More to our problem, one can imagine that, with the advent of the first digital quantum computers \cite{qc1}, the dynamics of quantum networks could be experimentally simulated, and, conversely, will have bearing on the behavior of running quantum algorithms. In particular, one has in mind population transfer and similar algorithms \cite{qc2,qc3,qc4} which can benefit from fast developement of ergodicity over a subset of preferred configurations (the equilibration time is the running time of the algorithm).

One of the fundamental aspects of quantum networks is the presence,
creation and transmission of entanglement. For qubits 
entanglement can be formed in different ways and one of them is the
use of CNOT operations between them. When such operations
are applied in networks the CNOT operations compete
against each other in entanglement formation (monogamy of
two-particle entanglement) and the asymptotic regime (state) of the
network is a density matrix of a rather simple form. The process can also 
be viewed as a competition between entanglement creation among the
qubits and decoherence originating from the imperfect control over the
system.

While the determination of the asymptotic regime is given by the
solution of a well-defined set of conditions specified by the
underlying graph structure (and independent of the actual weights of
the edges), the rates of convergence to the asymptotic is crucially
dependent of these weight. The structure of the
asymptotic state is quite clear and even accessible to analytic
treatment; on the other hand, the question of the convergence rates is largely 
unexplored and depends on a much larger set of parameters than the
structure of the asymptotic space. 

In the following we provide quantitative estimates for the convergence of a fully
connected graph undergoing CNOT interactions.
This simple example enables us both to demonstrate the influence of the network size (number of qbits) on the convergence rate at leading order, and discuss the influence of the geometry of the network, focusing in particular on two limiting cases, the complete graph and the circle graph. We shall also give a few hints on the influence of altering the edge probability (via the introduction of noise) on the convergency rate. 
A full understanding of our numerical outcomes will be obtained in the light of a general theorem on the convergency rate.

The properties to be discussed in the following sections make CNOT gates particularly appealing on a number of grounds.
First of all, they are entanglement forming (although, as discussed before, the map will induce dephasing, adding an  inherent competition in the global dynamics); second, they are mathematically simple, and in particular they form a finite group (see Secs. \ref{sec-observe} and \ref{sec-exex}); third, propagation and/or entanglement loss in an $n$-partite system are of general interest for the quantum information community; finally, CNOT is a special case of controlled rotations, which are the building blocks of dephasing in collisional models \cite{NAJ}.

This article is organized as follows. We set up the problem and
introduce notation in Sec.\ \ref{sec_setup}. In Sec.
\ref{sec-observe} we make an observation that leads to a drastic
simplification of the problem. In Sec. \ref{sec-exex} we clarify the
general framework in terms of a rather elementary example that can
be solved explicitly. The problem is then recast in simpler terms in
Sec.\ \ref{sec-simplerpr}. We perform a numerical analysis in Sec.\
\ref{sec-numerics}, and conclude in Sec.\ \ref{sec-concl}.

\section{Setting up the problem}
\label{sec_setup}

We consider $N$ qubits with the Hilbert space ${\cal H}=(\mathbb{C}^2)^{\otimes N}$ undergoing the following iterated dynamical evolution \cite{Holevo,NAJ}
\begin{eqnarray}
\Phi (\rho) &=& \sum_{i\in I} p_i U_i \rho U_i^{\dagger} ,
\label{dyn}
\end{eqnarray}
where state $\rho \in {\cal T}({\cal H})$ is a trace class operator on the
Hilbert space ${\cal H}$, $U_i$ are unitaries, $p_i$ a probability distribution and $I$ a set of indices (each representing a couple of qubits).
We shall focus on the behavior of the iterated map  $\Phi^n
(\rho) = \Phi (\Phi (\cdots (\Phi (\rho))))$ and its speed of convergence to equilibrium
when the $U_i$'s are CNOT gates, acting on qubits $a$ and $b$ according to
\begin{equation}
\label{cnotg}
U_{i=(a,b)}= \textrm{CNOT}_{(a,b)} \otimes \openone_{\textrm{rest}},
\end{equation}
and the interaction graph is fully connected. 
However, some of the ideas presented here are valid for more general networks and will be presented elsewhere.

\begin{figure}[h]
\centering
\includegraphics[width=.45 \textwidth]{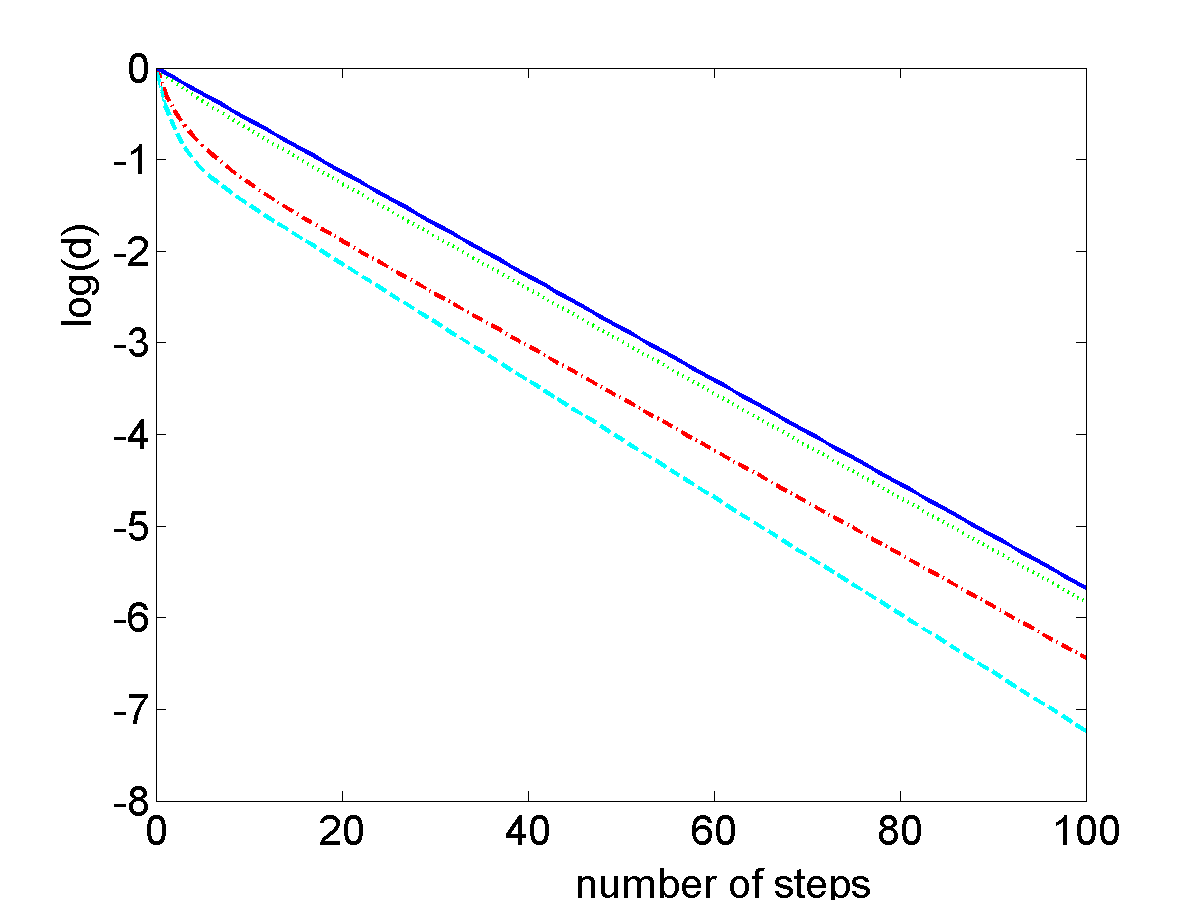}
\caption{Distance $d$ in Eq.\ (\ref{betastar}) between evolving states and their corresponding asymptotic limits: We considered a fully connected network of $6$ qubits with equally distributed weights. The (blue) solid line represents the upper bound in Eq.\ (\ref{betastar}). 
The (green) dotted, (red) dash-dotted and (light-blue) dashed lines correspond to initial states $|000001\>,|101010\>, |111111\>$, respectively.}
\label{convergence_fully_connected}
\end{figure}

We first observe that the superoperator (\ref{dyn}) is linear, so that its eigenvalues are Lyapunov coefficients.
Moreover, it is also normal, so that it admits an orthonormal set of attractors, spanning a linear manifold $\Pi$, and an additional orthonormal set corresponding to eigenvalues
$|\beta|<1$, the two sets making up a basis. Let $\rho(n)=\Phi^{n}(\rho(0))$ be the state after $n$-th iteration. Employing a Hilbert-Schmidt distance
$d(\rho_1,\rho_2)=||\rho_1-\rho_2|| =\left[\Tr(\rho_1-\rho_2)^2\right]^{1/2}$ one can prove that
\begin{equation}
\label{betastar}
d(\rho(n),\Pi) \leq (\beta_{*})^{n} d(\rho(0),\Pi) ,
\end{equation}
where $\beta_*= \max_{|\beta|<1} |\beta|$ denotes the maximal absolute value of the eigenvalues $|\beta|<1$ and is known as the \emph{subleading} eigenvalue of the map \cite{Fiedler1973,MoharDiameter}.
Hence, the speed of convergence to equilibrium is bounded by the subleading eigenvalue $\beta_{*}$.
A concrete example is shown in Fig.\ \ref{convergence_fully_connected}: one notices that the dynamics significantly depends on the initial state and the bound (\ref{betastar}) appears to be rather loose.

The aim of this article is to analyze the rate of convergence for (rather large) networks. This is a difficult problem, because it involves the eigenvalue of large superoperators (e.g., for as few as $20$ qubits the dimension of the superoperator is $2^{40} \simeq 10^{12}$) and depends on the rich structure of the interaction graphs. We shall therefore look for upper bounds of the subleading eigenvalue.

\section{A preliminary observation}
\label{sec-observe}

We start from an observation. Consider the random unitary channel (\ref{dyn})
\begin{equation}
\label{superoperator}
\Phi\left(\cdot\right)=\sum_{i \in I} p_i U_i \left( \cdot \right) U_i^{\dagger}
\end{equation}
and let $|e_i\>$ be an orthonormal basis of the Hilbert space $\hil H$. This map acts on the matrices $\rho$. 
In the basis $|e_i\rangle\langle e_j|$, since
\begin{eqnarray}
\textrm{Tr}( |e_b \rangle\langle e_a|\Phi(|e_i\rangle\langle e_j|))&=&\langle e_a| \Phi(|e_i\rangle\langle e_j|)|e_b\rangle\nonumber \\
&=&\sum_{i \in I} p_i \langle e_a|U_i| e_i\rangle(\langle e_b| U|e_j\rangle)^*
\nonumber \\
\end{eqnarray}
the matrix form of map $\Phi$ reads
\begin{equation}
\label{vector_form}
\Phi=\sum_{i \in I} p_i U_i \otimes U_i^{*},
\end{equation}
where $^*$ means complex conjugation. For the particular case (\ref{cnotg}) (CNOT gates, on which we shall focus)
\begin{equation}
U_i=U_i^\dag=U_i^*, \quad
{\rm and\ so\ }\quad  U_i^2=1,
\end{equation}
and we are therefore interested in the eigenvalues of the map
\begin{equation}
\Phi=\sum_{i\in I} p_i U_i\otimes U_i .
\end{equation}
Consider the operators $\cU_i=U_i\otimes U_i$ on the space $\cH\otimes \cH$. Notice that the group properties of $\cU_i$ and $U_i$ are exactly the same: $\cU_i^2=1$, while algebraic properties are not necessarily maintained, e.g.\ an equation $U_i U_j=a U_k$ would map to $\cU_i \cU_j=a^2 \cU_k$. However, for CNOT gates these equations never generate coefficients $a\neq 1$, so the full algebraic properties are maintained.

Notice also that $\Phi$ can be viewed as the average of the random process
\begin{equation}
F=\cU_i, \ \mathrm{with\ probability \;} p_i,
\end{equation}
in the sense that
\begin{equation}
\Phi=\avg{F}.
\end{equation}
Moreover, using the superoperator space trace,
\begin{equation}
\Tr\Phi=\Tr\avg{F}=\avg{\Tr F},
\end{equation}
and higher traces are connected to the multiplicative random process (MRP)
\begin{equation}
\Tr\Phi^n=\Tr\avg{\prod_{a=1}^n F_a} ,
\end{equation}
where $F_a$ are independent $F$ random variables. This is due irrespective of the definition of $\Tr$ but only to its linearity.

Using now that fact that for CNOT gates $\cU_i\sim U_i$ is an algebra-preserving isomorphism, it is not difficult to convince oneself that we obtain, for purpose of computing $\Tr(\Phi^n)$ and therefore the maximum Lyapunov exponent, a completely equivalent problem if we simplify the situation and consider the map
\begin{equation}
\phi=\sum_{i\in I} p_i U_i
\label{smallmap}
\end{equation}
and the random process
\begin{equation}
f=U_i\mathrm{\ with\ prob.\ } p_i,
\end{equation}
with associated MRP.
The usual definition of operator trace is used. Notice that some quantitative features comparing $F$ and $f$ are lost, since the values of the superoperator and operator traces are different. This gives rise to different spectra, although the important eigenvalues (the largest and second largest) are the same in all cases we have analyzed.

Let us now concentrate on the case of $f$ and let $G$ be the multiplicative group generated by the operators $U_i$. We write $G=\{ g_a\}_{a=1,...,M}$ and $1,U_i\in G$. In the case of the CNOT gates to be considered in this article, this group is a finite subgroup of the (finite) group $GL_n(\mathbf{F}_q)$.

Then
\begin{equation}
\Tr\phi^n=\sum_a k_a(n)\Tr(g_a),
\end{equation}
where $k_a(n)$ are coefficients measuring the probability that the MRP starting at $n=0$ in the identity, ends up in $g_a$ after $n$ steps. The MRP on the group is represented by a $M$-by-$M$ matrix $W_{a,b}$. The matrix is real and stochastic. Assuming it is also symmetric we can solve this problem using eigenvalues $\omega_a$ and eigenvectors $\ket{\omega_a}$ of the matrix $W$
\begin{equation}
k_a(n)=\sum_{b=1,...,M}\omega_b^n \braket{g_a}{\omega_b}\braket{\omega_b}{1},
\end{equation}
where $\ket{1}$ is the vector associated with the group element identity $g_a=1$.

\section{An example}
\label{sec-exex}

Let us look at an explicit example. Consider the case of two qubits with $p_1=p$ and $p_2=1-p$. The group $G=\{1,U_1,U_2,U_1 U_2, U_2U_1,U_1U_2U_1=U_2U_1U_2\}$. The matrix representing the MRP is
\begin{equation}
\label{wmap}
W=\left(
\begin{array}{cccccc}
 0 & p & 1-p & 0 & 0 & 0 \\
 p & 0 & 0 & 0 & 1-p & 0 \\
 1-p & 0 & 0 & p & 0 & 0 \\
 0 & 0 & p & 0 & 0 & 1-p \\
 0 & 1-p & 0 & 0 & 0 & p \\
 0 & 0 & 0 & 1-p & p & 0
\end{array}
\right),
\end{equation}
with eigenvalues
\begin{eqnarray}
\omega_a &=& \left(1,-1,-\sqrt{1-3 p+3 p^2},-\sqrt{1-3 p+3 p^2}, \right. \nonumber  \\
& & \left. \sqrt{1-3 p+3 p^2},\sqrt{1-3 p+3 p^2}\right).
\end{eqnarray}
The Perron-Frobenius theorem \cite{Graham}, together with stochasticity, guarantees that the maximum eigenvalue is $1$ and corresponds to the uniform eigenvector
\begin{equation}
\braket{g_a}{\omega_1}=\frac{1}{\sqrt{6}}.
\end{equation}
The eigenvector corresponding to eigenvalue $-1$ is
\begin{equation}
\braket{g_a}{\omega_2}=\frac{(-1)^{P_a}}{\sqrt{6}}.
\end{equation}
where $P_a$ is the parity of $g_a$, which is $0$ for $1,U_1U_2,U_2U_1$ and $-1$ for the remaining 3 elements. The remaining subleading eigenvalues $s=\pm\sqrt{1-3p+3p^2}$ with their associated eigenvectors $\ket{s_c}_{c=1,..,4}$.
Notice that the subleading eigenvalue coincides with that obtained by explicit calculation.
Putting all together we find
\begin{equation}
\Tr\phi^n=\frac{1}{6}1^n\sum_a\Tr(g_a)+\frac{1}{6}(-1)^n\sum_a(-1)^{P_a}\Tr(g_a)+s^n A.
\end{equation}
Notice that
\begin{eqnarray}
\sum_a\Tr(g_a)&=&12,\\
\sum_a(-1)^{P_a}\Tr(g_a)&=&4+1+1-2-2-2=0,
\end{eqnarray}
so that
\begin{equation}
\Tr\phi^n=2+s^n A,
\end{equation}
which means that the leading and sub-leading eigenvalues of $\phi$ are given by the leading and sub-leading eigenvalues of $W$. Notice how the $\Tr(g_a)$ is reflected in the degeneracy of the eigenvalues (for example 2 for the eigenvalue 1).

The only non-trivial step is the cancellation of the contribution of the eigenvalue $-1$. It is not difficult to prove that i) this eigenvalue always exists and it is due to the fact that parity breaks the group $G$ in two, $G_+$ and $G_-$ and that the MRP necessarily connects $G_\pm\to G_\mp$, and ii) that its contribution, once the trace is taken, is always 0.
Therefore, we can assert that the sub-leading eigenvalue of $W$ is the sub-leading eigenvalue of $\phi$.

We now go back to the problem of the map $\Phi$. One can repeat exactly the same steps as before if we identify the group $G$ as generated by $U_1\otimes U_1$ and $U_2\otimes U_2$. This is exactly the same group $G$ as before (it is the diagonal projection of the group $G\otimes G$) and therefore the very same calculations occur. There is one, crucial difference as the superoperator trace $\Tr$ will give different results from the operator trace $\Tr$. This affects the polynomials $\Tr{\phi^n}$ and $\Tr{\Phi^n}$ but we observe that this \emph{does not change the subleading eigenvalue} which is the same in both maps in all the examples we have checked. Therefore, the rate to approach to ergodicity is the same for both maps.

\section{A simpler problem and some bounds}
\label{sec-simplerpr}

The observation in Sec.\ \ref{sec-observe} and the explicit example in Sec.\ \ref{sec-exex} show that the eigenvalue problem can be significantly simplified for certain types of interaction graphs and unitaries. Consider the random unitary channel $\Phi$ in Eq.\ (\ref{dyn}) and let $\sigma(\Phi)$ be the spectrum of $\Phi$ and $\sigma_1(\Phi)$ the set of the elements of $\sigma(\Phi)$ with magnitude one. Consider now the operator
\begin{equation}
\label{simple_operator}
\phi=\sum_{i \in I} p_i U_i.
\end{equation}
Then for certain quantum networks the subleading eigenvalue of the operator (\ref{vector_form}) coincides with the subleading eigenvalue of the operator (\ref{simple_operator}) and both are positive, i.e.
\begin{eqnarray}
\label{subleading_rule}
\beta_*(\Phi) = \sup_{\lambda \in \sigma(\Phi) \setminus \sigma_1(\Phi)} \lambda = \beta_*(\phi) = \sup_{\lambda \in \sigma(\phi) \setminus \sigma_1(\phi)} \lambda.
\end{eqnarray}
This property has been numerically checked and appears to be valid for a wide class of \emph{fully connected} graphs (made up of e.g., unitary transpositions, controlled and/or local rotations and the special case of CNOT gates analyzed in this article). On the other hand, it is not valid for general (e.g., non fully connected) graphs. We shall assume henceforth that the property is valid, at least in the cases to be investigated in this article.

\begin{figure}[t]
\includegraphics[width=6.cm]{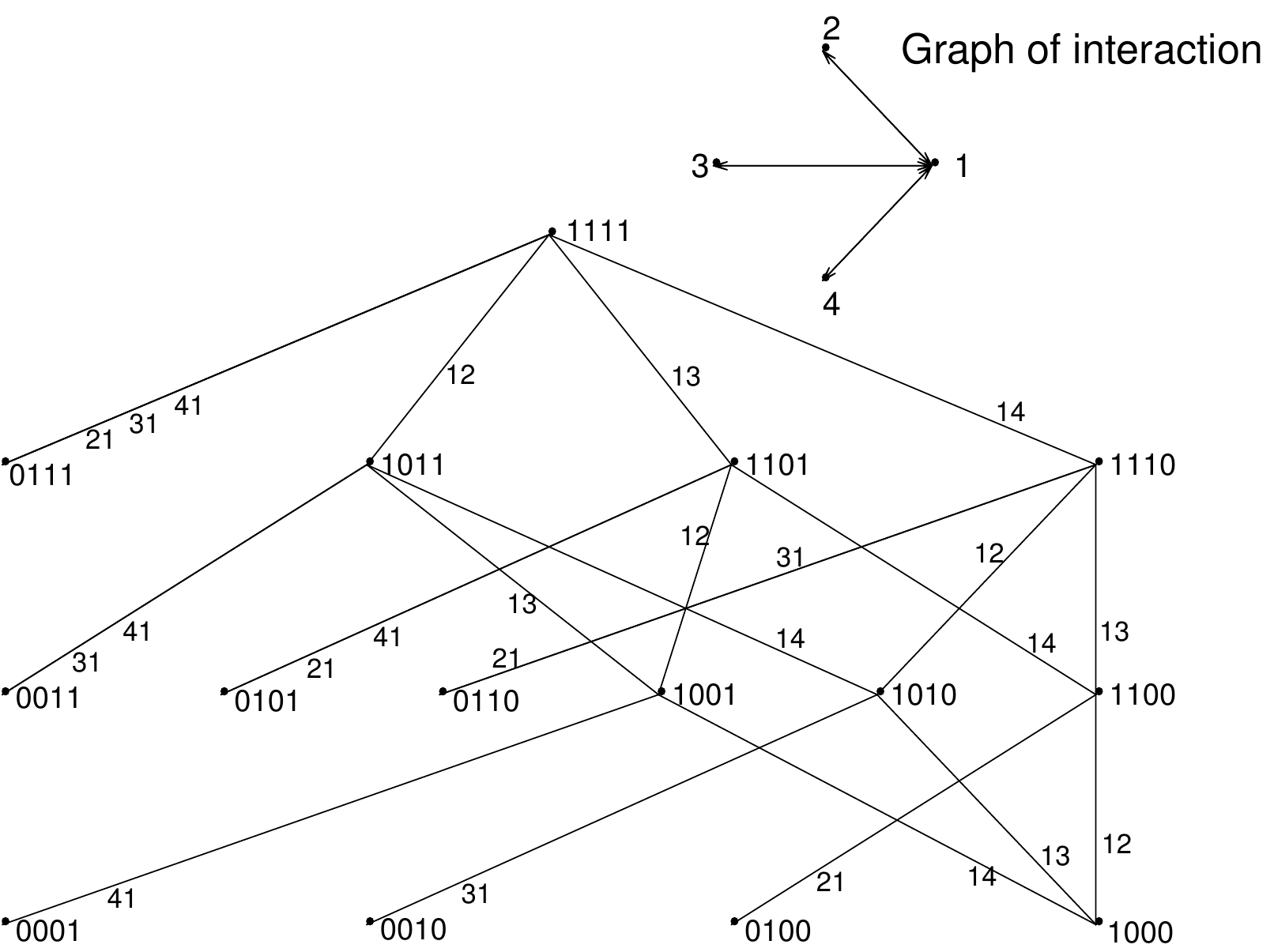}
\caption{Graph $\set G_{\phi}$ induced by graph of interaction: 4-qubit oriented star. Vertices correspond to elements of computational basis. Two vertices are linked iff one of them is an image of the other under some CNOT gate from the interaction graph. The list of pairs of indices are the CNOT gates which contribute to weights of a given edge.}
\label{graph3}
\end{figure}

We now turn to the study of the specific features of the subleading eigenvalue of operator (\ref{simple_operator}) when all unitaries $U_i$ are CNOT gates acting on two qubits only, as in (\ref{dyn})-(\ref{cnotg}), and leaving the remaining ones unchanged. In such a case, in the computational basis, operator (\ref{simple_operator}) is bistochastic or doubly stochastic
\begin{equation}
\sum_{i=1}^{2^N}\phi_{ij}=\sum_{j=1}^{2^N}\phi_{ij}=1, \quad \forall \; 1 \leq i,j \leq 2^N
\end{equation}
and is the adjacency matrix of an undirected weighted graph $\set G$, whose vertices are elements of the computational base and where there is an edge between two vertices (basis vectors) whenever one vector is the image of the second one under application of some $U_i$. Due to this definition the graph can contain loops.
The weight of an edge is the sum of the probabilities assigned to the unitary operations that define this edge. In other words, a unitary operator $U_i$ contributes to the weight of an edge with probability $p_i$ if the vectors corresponding to the end vertices of the edge are images of each other under application of this $U_i$
(edges are not directed, as $\textrm{CNOT}^2 = \openone$). This definition applies also for loops.
As a simple consequence, the weights of the edges adjacent to a given vertex, with the inclusion of the weights of loops, always sum up to one. Now, the operator $\phi$ is simply the adjacency matrix of this weighted graph $\set G$.
Because the interaction graph is fully connected, the graph $\set G$ has two components of continuity: the vertex corresponding to the vector with zero excitation and the remaining connected vertices. The one-vertex component contributes to the spectrum of Laplacian matrix by one zero eigenvalue.
Let us remove this vertex from the graph in order to still have a connected graph $\set G_{\phi}$ with its adjacency matrix $A_{\phi}$ and its Laplacian matrix given by \cite{Chung1997,Mohar1991}
\begin{equation}
\label{laplacian_cnot}
L_{\phi}=\openone-A_{\phi},
\end{equation}
where $\openone$ is the $(2^N-1)\times (2^N-1)$ identity matrix.
Equation (\ref{laplacian_cnot}) follows from the fact that the degree of each vertex is one.
An example of such a graph $\set G_{\phi}$ is displayed in Fig. \ref{graph3}. The interaction graph is the 3-qubit oriented star. Only edges contributing to the weights of a given edge are explicitly displayed. Loops are not displayed.

Let us denote the algebraic connectivity \cite{Abreu2007} (the second-smallest eigenvalue (counting multiple eigenvalues separately) of the Laplacian matrix of G)
of the graph $\set G_{\phi}$ by $\gamma_{\phi}$. From the previous discussion it follows that
\begin{equation}
\label{subbeta}
\beta_{*}(\Phi)=\beta_{*}({\phi})=1-\gamma_{\phi}
\end{equation}
Hence, spectral graph theory can be employed to find a good
lower bound for the algebraic connectivity and use it to upper bound the subleading eigenvalue of the superoperator $\Phi$.

In order to obtain meaningful bounds, let us start by noting that
\begin{equation}
\label{bound_generalized}
\gamma_{\phi} \geq \frac{4}{(2^N-1) \textrm{diam}(\set G_{\phi})},
\end{equation}
``diam" being the diameter of the graph.
This is easy to prove \footnote{There are better bounds, but all of them depend on the graph parameters, whose determination is at least as difficult as the determination of the spectrum.}. This bound has the advantage of being valid for an arbitrary interaction graph with an arbitrary probability distribution. The disadvantage is that it exponentially depends on the number $N$ of qubits.
Of course, this simply recasts the problem in terms of determining
the diameter of the (weighted or unweighted) graph $\set G_{\phi}$ (which is possible at least in particular cases).
For example, for a circle unweighted interaction graph with $N$ vertices, the associated graph $G_{\phi}$ has $\textrm{diam}(\set G_{\phi})=2(N-1)$, while for a fully connected unweighted graph with $N$ vertices, the associated graph $G_{\phi}$ has $\textrm{diam}(\set G_{\phi})=N$. The diameter of weighted graphs are much more involved.

As shown Fig.\ \ref{graph3}, the vertices corresponding to the same number of excitations are not connected. They can be connected only to vertices whose number of excitation differ by one \footnote{More precisely, two vertices can be connected only if they differ in the excitation of one qubit. Whether they are really connected then depends on the details of the interaction graph.}. 

In the case we are considering (fully connected interaction graph with random CNOT gates acting on pairs of qubits), two vertices of the graph $\set G_{\phi}$ are connected if and only if their excitations differ only for one qubit. Therefore, omitting the weights of the graph $\set G_{\phi}$, the associated unweighted graph $\set G_{\phi}^{'}$ is almost the hypercube $\set H_n$, the only difference being that the graph $\set G_{\phi}^{'}$ does not contain the vertex $|0,0,\ldots,0\>$ and its adjacent edges.  As a consequence \cite{Abreu2007},
$\gamma_{\phi}^{'} \geq \gamma_{H_n}-1 = 2-1=1$
and therefore
\begin{equation}
\label{bound_fully}
\gamma_{\phi} \geq \min_{ij \in I} p_{ij}.
\end{equation}
This is the bound shown in Fig.\ \ref{convergence_fully_connected}:
it depends very weakly on the weights, but does not depend exponentially on the number of qubits, which makes it useful.
A relevant situation is when the probabilities are equally distributed on the $N$ (fully connected) qubits
\begin{equation}
\label{eqdistr}
p_{ij}=\frac{1}{N(N-1)}, \quad \forall \; i,j .
\end{equation}

\section{Numerical analysis}
\label{sec-numerics}

We now turn to a numerical analysis. The analysis has been performed on the CRESCO/ENEAGRID High Performance Computing infrastructure \cite{enea}.
The numerical evaluations have been performed on the ``small" map $\phi$ in Eq.\ (\ref{smallmap}), for
fully connected graphs of CNOT gates (\ref{cnotg}). In the light of our discussion, the results are also valid for the ``larger" map $\Phi$ in Eq.\ (\ref{dyn}).
For the sake of comparison, we also performed some analyses for interaction \emph{circle} graphs of CNOT gates.

\begin{figure}[t]
\includegraphics[width=8.5cm]{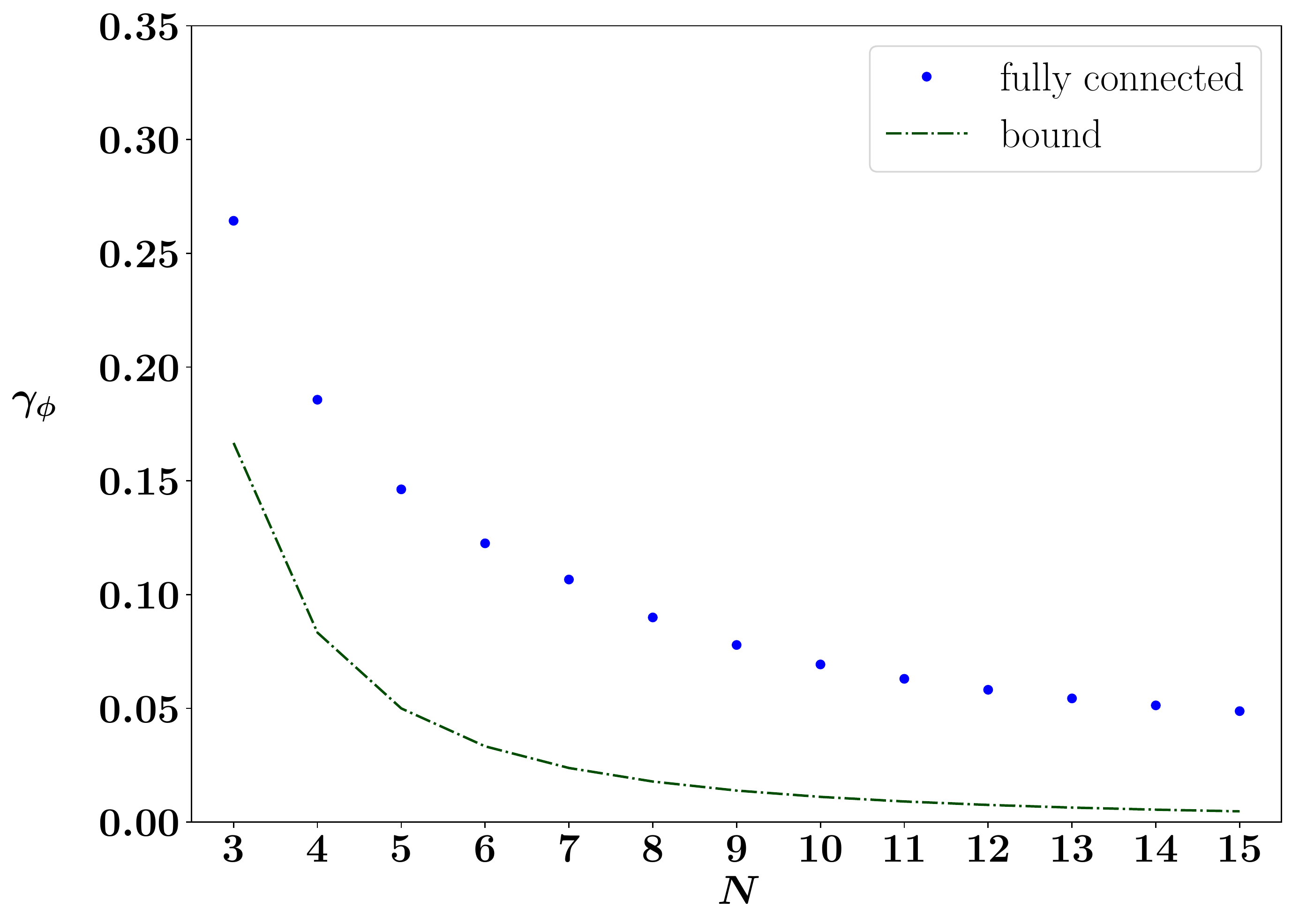}
\caption{Fully connected graph with equally distributed weights. Dots: algebraic connectivity. Dash-dotted (green) line: bound (\ref{bound_fully}).
}
\label{graph_connectivity}
\end{figure}

Figure \ref{graph_connectivity} displays the algebraic connectivity $\gamma_{\phi}$ of a fully connected graph with equally distributed weights, as in (\ref{eqdistr}), for $3 \leq N \leq 15$. The bound (\ref{bound_fully}) is shown for comparison and appears to be far from tight. The approach to equilibrium, measured by
the subleading eigenvalue of the superoperator $\Phi$, $\beta_{*}(\Phi)=\beta_{*}({\phi})$ in Eq.\ (\ref{subbeta}), is slower for increasing $N$, as intuitively expected.
We now try to unveil the $N$ dependence.

\begin{figure}[t]
\includegraphics[width=8.5cm]{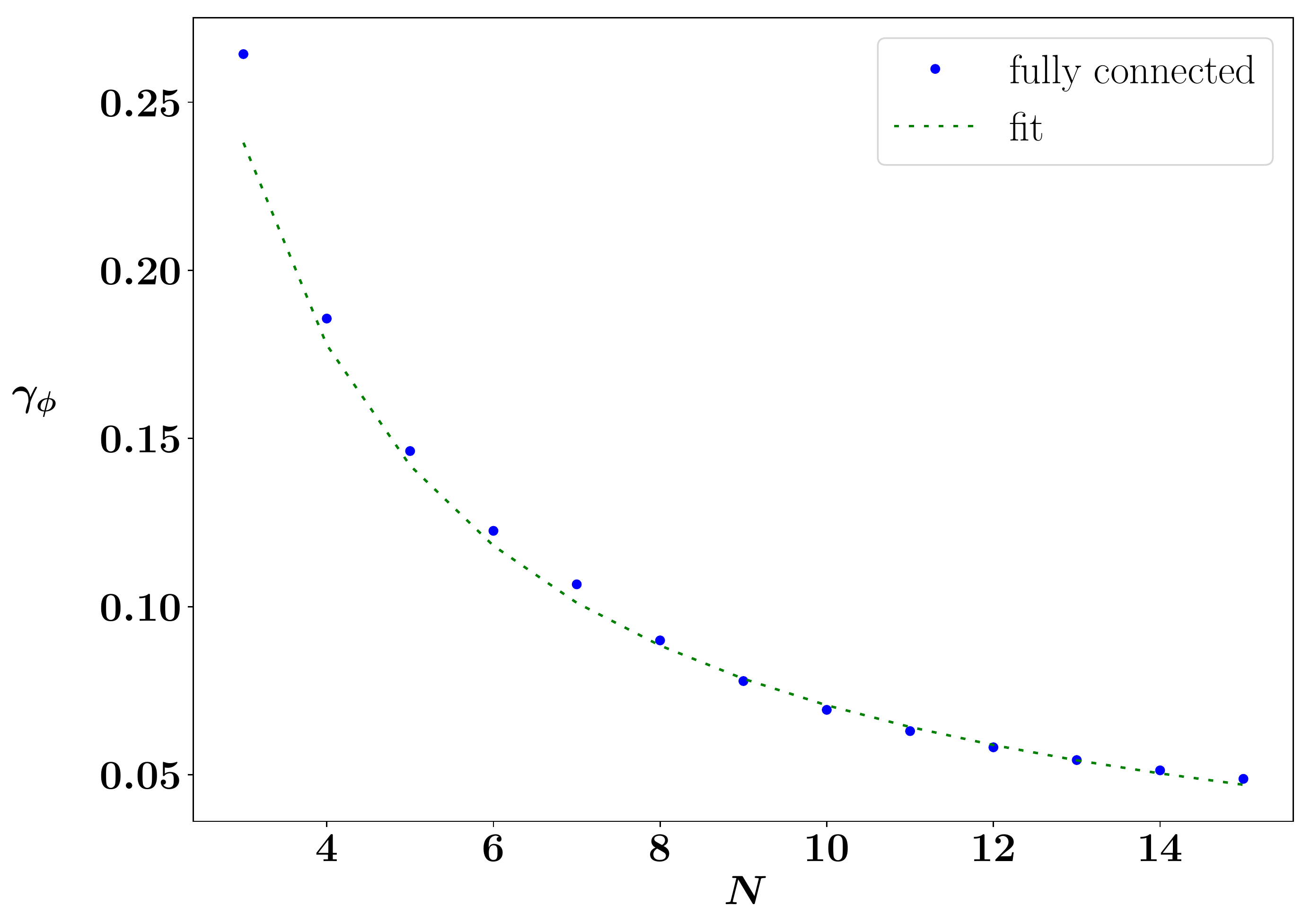}
\caption{
Dots: algebraic connectivity. Dotted (green) line, fit in Eq.\ (\ref{fitpl2}), with  $a = 0.704$ and $b= 0.030$, obtained by fitting the points $N \geq 8$. }
\label{graph_connectivitylog}
\end{figure}

The function $\gamma_{\phi} = a/N$, with $a = 0.707$ yields a good fit. As can be seen in Fig.\ \ref{graph_connectivitylog}, the addition of a contribution $O(N^{-2})$ yields an excellent fit at large $N$:
\begin{eqnarray}
\gamma_{\phi} = \frac{a}{N} + \frac{b}{N^2} ,
\label{fitpl2}
\end{eqnarray}
with $a = 0.704$ and $b= 0.030$. We emphasize that other functional forms (such as different power laws) do not yield equally good results. We offer no explanation for the $N$-dependence in Eq.\ (\ref{fitpl2}).

\begin{figure}[h]
\includegraphics[width=8.5cm]{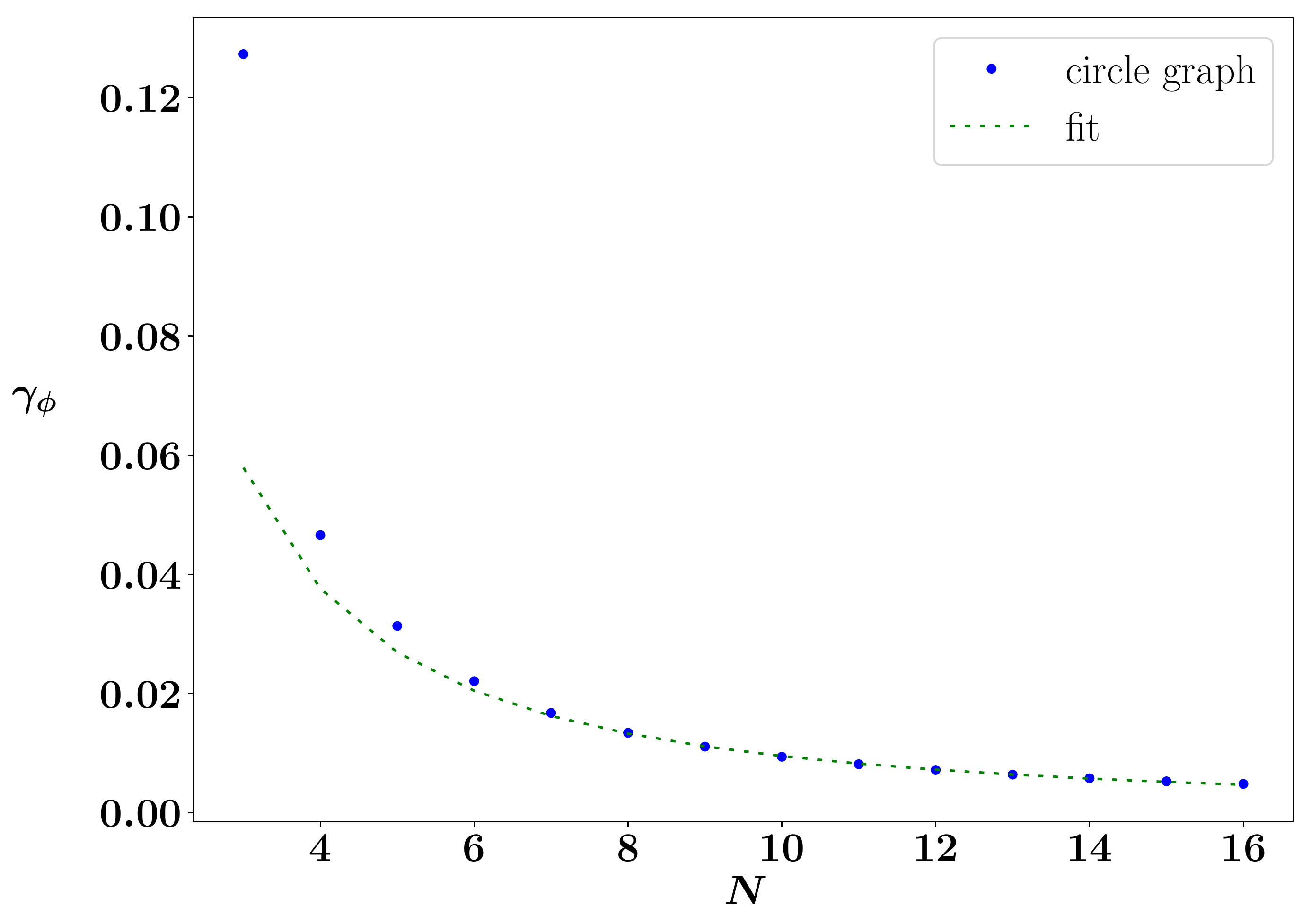}
\caption{Same as in Fig.\ \ref{graph_connectivitylog}, for the circle graph.
Dots: algebraic connectivity. Dotted (green) line: fit in Eq.\ (\ref{fitcircle}), with $a=0.301$ and $b = -1.888 \times 10^{-4}$, obtained by fitting the points $N\geq 8$.}
\label{log_graph_connectivity}
\end{figure}

In Fig.\ \ref{log_graph_connectivity} we show the $N$-dependence of the algebraic connectivity for a circle graph with equally distributed weights. This yields a good comparison with the data displayed in Figs.\ \ref{graph_connectivity}-\ref{graph_connectivitylog}. The approach to equilibrium is much slower, as was to be, being the graph less connected. The best fit yields the functional dependence
\begin{eqnarray}
\gamma_{\phi} = \frac{a}{N^{3/2}} + \frac{b}{N^{5/2}} ,
\label{fitcircle}
\end{eqnarray}
with $a=0.301$ and $b = -0.189 \times 10^{-3}$.

\begin{figure}[h]
\includegraphics[width=8.5cm]{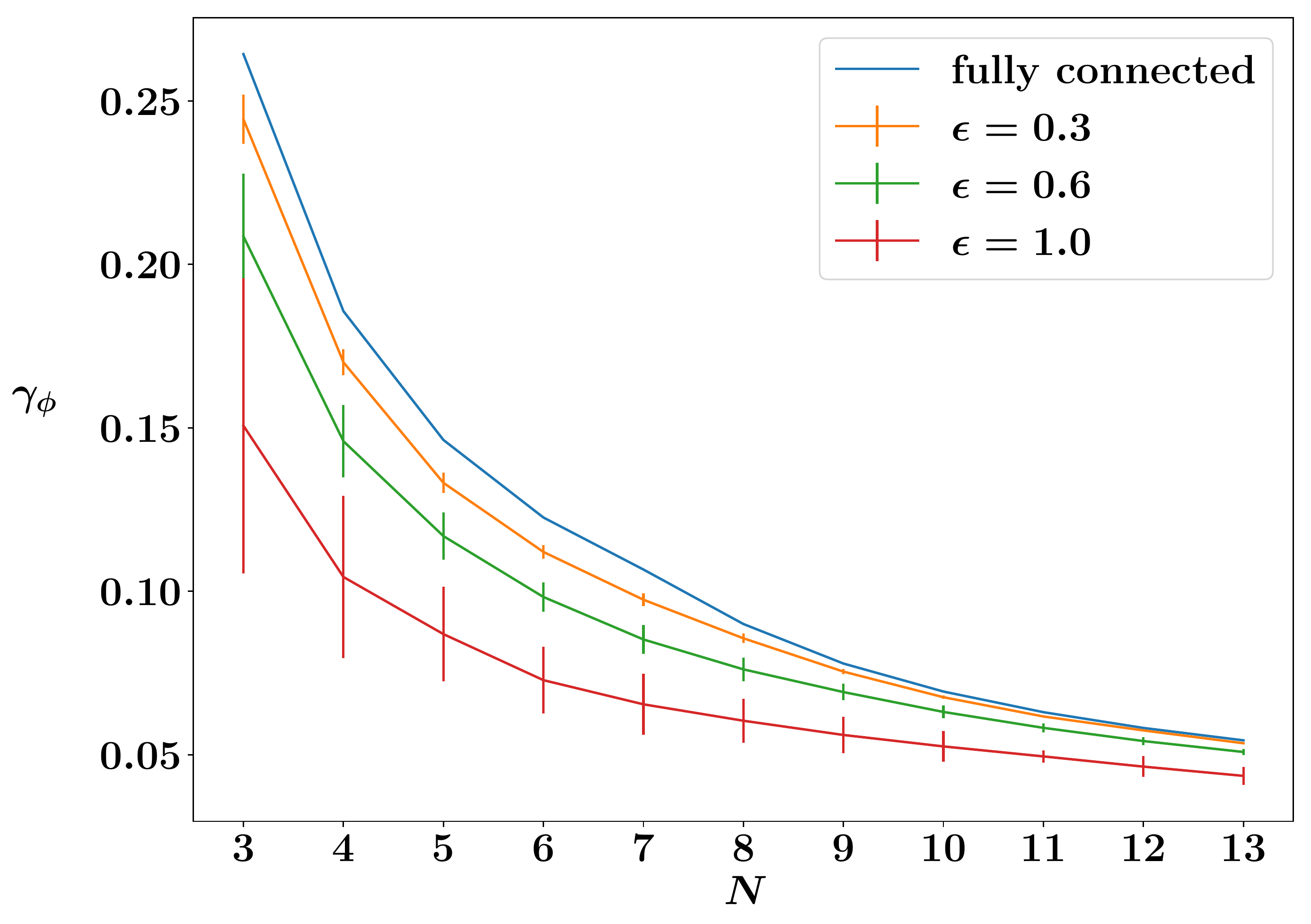}
\caption{Algebraic connectivity of a fully connected graph vs $N$, for different noise on the probabilistic weights $p_{ij}$. Noise $\epsilon$ increases from top to bottom. The vertical bars are standard deviations. The fits (not shown) always yield a dependence $\gamma_{\phi}= a/N + b/N^2$, with
($\epsilon = 0.3$) $a= 0.695,  b=-0.117$;
($\epsilon = 0.6$) $a= 0.735, b= -1.013$;
($\epsilon = 1.0$) $a=0.702, b= -1.763$.
}
\label{boxprob_fully}
\end{figure}

\begin{figure}[h]
\includegraphics[width=8.5cm]{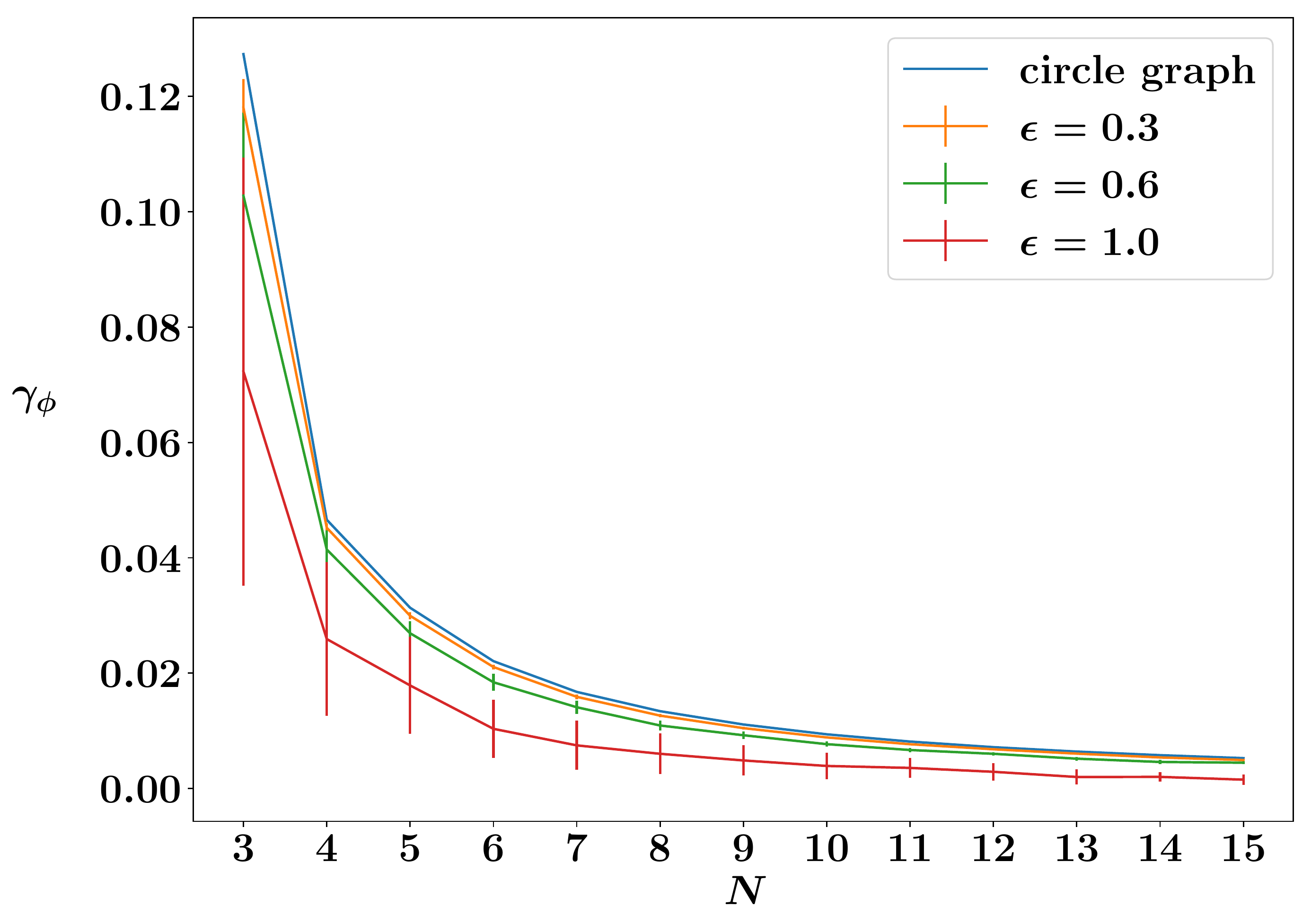}
\caption{Same as in Figure \ref{boxprob_fully}, for a circle graph. Noise $\epsilon$ increases from top to bottom. The vertical bars are standard deviations.}
\label{boxprob_circle}
\end{figure}

We now turn to an analysis of the convergence features of the network in the presence of noise. This will yields additional insight into the mechanisms of convergence and the bound
(\ref{bound_fully}).

In Fig.\ \ref{boxprob_fully}, we add a noise to the probabilistic weights in Eq.\ (\ref{eqdistr}). Each probability is multiplied by a random number in the interval $[1-\epsilon, 1+\epsilon]$; the probabilities are subsequently normalized so that they sum up to 1. We take $\epsilon=0.3, 0.6, 1$ (increasing noise), the last figure being the maximum value if the positivity of the probabilities is to be preserved. The presence of noise yields a slower approach to equilibrium. This result, at first a bit surprising, is understood by realizing that noise makes some probabilities smaller, so that some nodes are more isolated than others, and tend to equilibrate later. Notice that, at a given noise realization, the iterated map is always the same. Observe also that this behavior is in qualitative accord with the philosophy behind the bound
(\ref{bound_fully}).

In Fig.\  \ref{boxprob_circle} we add a noise to the weights of a circle graph. As expected, in the light of the preceding comments, some nodes become more isolated than others, and the approach to equilibrium is significantly slower (as, unlike with the fully connected graph, is it now easier to create more isolated nodes). This corroborates and completes the  picture discussed above.

One problem that remains to be understood is whether the bound (\ref{bound_fully}) can be saturated in some sense. A moment's reflection shows that, in general, this cannot be the case. Indeed, take one of the $p_{ij}=0$ in Eq.\ (\ref{bound_fully}); the graph is almost fully connected (only one link is missing), but will nonetheless tend to equilibrium, as there are many nonvanishing links between any given qubit and the other qubits in the network. For such a graph, $\gamma_{\phi}$ must be strictly positive and $\beta_*$ in Eq.\ (\ref{subbeta}) strictly smaller than 1.

\begin{figure}[t]
\includegraphics[width=8.5cm]{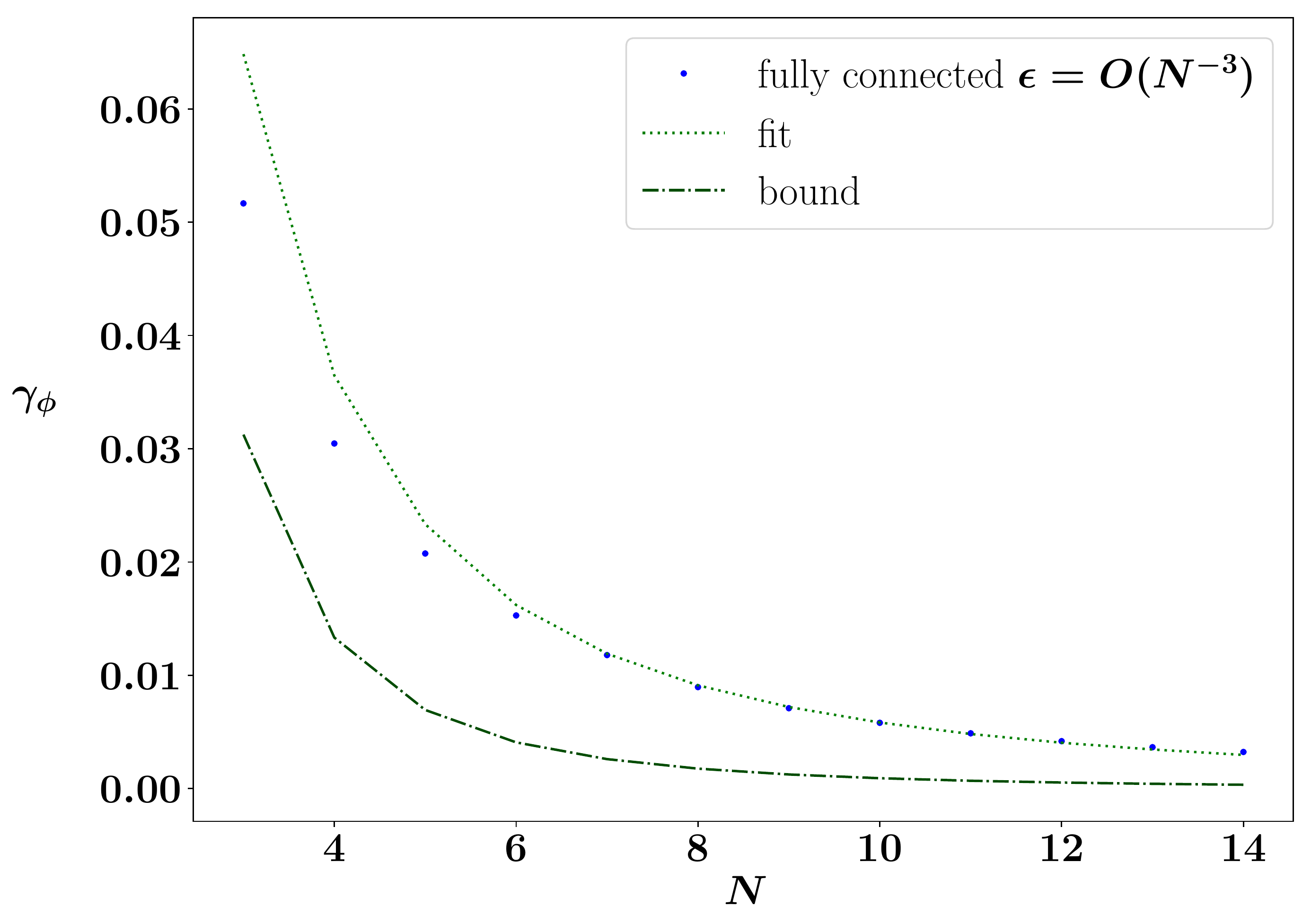}
\caption{A very unbalanced fully connected graph. $((N(N-1) - 1)$ links have probabilities $\epsilon = O(N^{-3})$, while one link has probability $O(1)$.
Dots: algebraic connectivity.
Full (green) line: fit in Eq.\ (\ref{smalleps}), with $a=0.629$ and $b = -3.831$, obtained by fitting the points $N\geq 8$.
Dash-dotted (green) line: bound (\ref{bound_fully}).
 }
\label{smalle}
\end{figure}

Motivated by the preceding comments, in Fig.\ \ref{smalle} we considered a fully connected network with very unbalanced weights: we took $((N(N-1) - 1)$ links with probability $\epsilon = O(N^{-3})$, \emph{except} one link with a probability $O(1)$, so that
\begin{equation}
\label{epsenne}
\min_{ij \in I} p_{ij} = \frac{N^{-3}}{(1 + ((N(N-1) - 1)N^{-3})} \sim \frac{1}{N^3}.
\end{equation}
As expected, in the light of the preceding comments and discussion, the approach to equilibrium is very slow (in fact, the slowest we have observed in our numerical simulations). However, as can be seen, the bound (\ref{bound_fully}) is not saturated.
A fit yields the dependence
\begin{eqnarray}
\gamma_{\phi} = \frac{a}{N^{2}} + \frac{b}{N^{4}} ,
\label{smalleps}
\end{eqnarray}
with $a=0.629$ and $b = -3.831$ (by fitting points with $N \geq 8$).

Summarizing, the numerical analysis shows that there are a number of factors that influence the approach to equilibrium in a fully connected quantum network of qubits whose interactions are represented by CNOT gates. Change of connectivity and weights can induce very different convergence rates, even at the qualitative level. The mathematical bound
(\ref{bound_fully}) is clearly valid, but appears to be loose in most situations. Further analyses are required to elucidate the underlying equilibration mechanisms.

\section{Conclusions}
\label{sec-concl} We studied the rate of convergence of full graph quantum networks. Using analytic methods we gave estimates for
this rate and by using numerical methods we determined the
rates of convergence to the asymptotic state. The convergence is inversely proportional to the number of
vertices-qubits forming the graph. The expansion coefficients have
been determined numerically. The numerical tests are limited up
to 15-16 qubits, which turn out to be sufficient to determine the convergence rates at leading order. 

The estimate of convergence rates is clearly of fundamental importance, as it
is one of the basic parameters characterizing random quantum networks. At the same time,
the convergence rate is also of practical importance, as it gives 
experimental physicists the typical scale after which the network
``equalizes" and its asymptotic is reached. As repeatedly mentioned, 
the determination of the asymptotics is in many cases
accessible by analytic methods and can be worked out in detail.
However, results (especially analytic ones) on the convergence rates
are scarce. Our estimates, both analytic and numerical, are a step into this uncharted territory. 

The observation in Sec.\ \ref{sec-observe} enabled us to
drastically simplify the problem, significantly reducing its complexity. 
As already emphasized, besides such inherent complexity, a number of factors (connectivity, topology, probabilistic weights) heavily influence the approach to equilibrium, yielding very different convergence rates.
While our work is limited to special types of graphs, we expect the
approach and results to hold also for similar situations. 
Further work is needed in order to scrutinize the underlying equilibration mechanisms and possibly generalize the results discussed in this article to different networks.

\section*{Acknowledgements}
SP would like to thank the Department of Physics of the Czech
Technical University in Prague for their warm hospitality. The
numerical analysis has been done on the CRESCO/ENEAGRID High
Performance Computing infrastructure of ENEA.
JN and IJ acknowledge the financial support from
RVO14000 and "Centre for Advanced Applied Sciences", Registry No.
CZ.02.1.01/0.0/0.0/16 019/0000778, supported by the Operational
Programme Research, Development and Education, co-financed by the
European Structural and Investment Funds and the state budget of the
Czech Republic and by the Czech Science foundation (GACR) project
number 16-09824S. IJ was partially supported from GACR 17-00844S.
SP acknowledges support by MIUR via PRIN 2017 (Progetto di Ricerca di Interesse Nazionale), project QUSHIP (2017SRNBRK), by INFN through the project ``QUANTUM" and by Regione Puglia and QuantERA ERA-NET Cofund in Quantum Technologies (GA No.\ 731473), project PACE-IN. The work of AS is within the activities of the TQT, Trieste.

\end{document}